\documentclass{PoS}

\usepackage{amsmath}
\usepackage{color}

\newcommand{\etal}{\textsl{et al.}}
\newcommand{\PR}{Phys.\ Rev.}
\newcommand{\PRL}{Phys.\ Rev.\ Lett.}

\newcommand{\PTP}{Progr.\ Theor.\ Phys.}

\newcommand{\peiii}{$\pi_{\text{e}3}$}
\newcommand{\td}{\text{d}}

\newdimen\arrayruleHwidth
\makeatletter
\def\Hline{\noalign{\ifnum0=`}\fi\hrule \@height \arrayruleHwidth
    \futurelet \@tempa\@xhline}
\makeatother
\title{Evaluating $V_{ud}$ from neutron beta decays}

\ShortTitle{$V_{ud}$ from neutron beta decays}

\author{\speaker{Dinko Po\v{c}ani\'c}\thanks{Work supported by grant
    PHY-1614839 from the United States National Science Foundation.}\\
        University of Virginia, Charlottesville, VA 22904-4714, USA\\
        E-mail: \email{pocanic@virginia.edu}}

\abstract{Although well studied, the neutron still offers a unique
  laboratory for precise tests of Standard Model (SM) predictions.
  Neutron decay is free of nuclear structure corrections present in
  nuclear beta decays, and, with a $10^8$ times larger branching ratio
  than the theoretically cleaner pion beta decay, it is more readily
  accessible to experimental study than the latter.  Measurements at
  sufficient precision of the neutron lifetime, and of correlations in
  free neutron beta decay, offer several stringent tests of the SM,
  including the weak quark couplings (quark-lepton universality), and
  certain extensions beyond the standard $V-A$ weak interaction
  theory. This paper focuses on the long-running free neutron beta decay
  experimental program aimed at obtaining an independent determination
  of the Cabibbo-Kobayashi-Maskawa (CKM) mixing matrix element $V_{ud}$.
  We discuss the present state of precision achieved in this program
  and briefly review the currently active projects, as well as the
  expected near-term improvements in the field.}

\FullConference{9th International Workshop on the CKM Unitarity Triangle\\
		28  November - 3 December 2016\\
		Tata Institute for Fundamental Research (TIFR), Mumbai, India}

\begin{document}

\section{Motivation}

In one form or another, quark-lepton universality has been a prominent
pillar of the evolving description of the fundamental interactions in
particle and nuclear physics since the 1950s.  Its systematic treatment
was completed with the introduction of the Cabibbo-Kobayashi-Maskawa
mixing matrix \cite{Cab63,Kob73} which summarizes the flavor-changing
processes induced by weak interactions of quarks.  Examining the
properties of the 3$\times$3 unitary CKM matrix in detail allows us to
address a number of questions of considerable interest to mapping the
limits of the present Standard Model.

The CKM matrix unitarity is one among these questions.  A departure from
unitarity would indicate the presence of new physics (a possible 4th
fermion generation, additional bosons, or compositeness in the known
ones, supersymmetric particles, leptoquarks, etc.).  The most precise
test of the CKM matrix unitarity is performed in its first row, i.e., by
evaluating
\begin{equation}
   \Delta = 1 - |V_{ud}|^2 - |V_{us}|^2 - |V_{ub}|^2\,,
\end{equation} 
where $|V_{ub}|^2\simeq 1\times 10^{-5}$ is negligible.  The most recent
global evaluation of the CKM matrix unitarity yields $\Delta =
0.0005(5)$, in excellent agreement with the 3-generation SM prediction
\cite{Blu16}.  Leaving $V_{us}$ aside, the $V_{ud}$ input is essentially
determined by the measurements of rates of several superallowed Fermi
(SAF) $0^+\to 0^+$ nuclear beta decays \cite{Har15}.  SAF decay
measurements are carried out with impressive precision; however their
interpretation is limited by theoretical uncertainties, which also
include the uncertainties related to nuclear structure corrections.
While the latter are claimed to be well controlled, an independent
experimental determination of $V_{ud}$ in systems not subject to nuclear
structure effects would be highly desirable.  Two such systems present
themselves:

Not involving any baryons, the pion beta decay,
$\pi^+\to\pi^0\,\text{e}^+ \nu_{\text{e}}$ or \peiii, a pure vector
$0^-\to 0^-$ transition, is the theoretically cleanest probe of
$V_{ud}$.  It is, however, a rare decay with a branching ratio of
$\mathcal{O}(10^{-8})$, which makes high precision measurements
extremely challenging.

Incorporating both vector and axial vector amplitudes, neutron beta
decay, n $\to$ p\,e$^-\,\bar{\nu}_{\text{e}}$, is more complex but
experimentally more accessible than \peiii\ decay.  This complexity, on
the other hand, provides redundant ways to measure the SM-predicted
observables, providing independent cross-checks of possible new physics,
including tensor interactions, supersymmetric as well as right-handed
extensions to SM, etc., with competitive sensitivity \cite{Bae14}.

We next turn our attention to precision studies of the neutron beta
decay with the express goal to extract the CKM matrix element $V_{ud}$.

\section{Basics of neutron beta decay}

\begin{figure}
  \centerline{
    \parbox{0.8\linewidth}{
      \includegraphics[width=\linewidth]{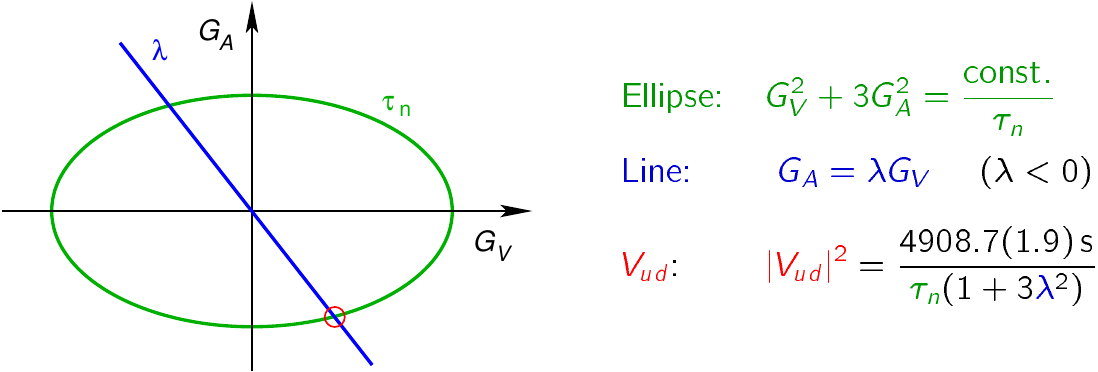}
                          }
              }
  \caption{Determination of $V_{ud}$ from neutron decay requires two
    independent measurements: decay correlations or asymmetry to
    evaluate the form factor ratio $\lambda=G_A/G_V$, and neutron
    lifetime or decay rate, $\tau_{\text{n}}^{-1}\propto G_V^2+3G_A^2$.  The
    evaluation of $|V_{ud}|$ is illustrated in the $(G_V,G_A)$ parameter
    plane as the intersection of an ellipse and a line, denoted by a
    small red circle.}
  \label{fig:Ga_Gv_ellipse}
\end{figure}

With a three-body final state, all beta decays are described by several
experimental observables.  Since only the vector amplitude is subject to
weak interaction universality through vector current conservation, the
axial vector contribution complicates the determination of $V_{ud}$ in
neutron decay.  Thus, unlike SAF and \peiii\ decays, measuring
$\tau_{\text{n}}$, the neutron lifetime (or decay rate), is not enough;
an additional observable, $\lambda=G_A/G_V$, the ratio of axial vector
to vector form factors, has to be measured, as illustrated in
Figure~\ref{fig:Ga_Gv_ellipse}.
Measurements of $\tau_{\text{n}}$ and of the ratio $\lambda$ each
present their own challenges, briefly examined below.  We close this
section with considerations of the experimental determinations of
$\lambda$.  The standard expression for the differential beta decay rate
for particles with spin is given by \cite{Jac57}:
\begin{equation}
  \frac{\td^5\Gamma}{\td E_{\text{e}}\,\td^2\!\Omega_{\text{e}}\,\td^2\!\Omega_\nu}
   \propto 
    p_{\text{e}} E_{\text{e}} \left( E_0-E_{\text{e}}\right)^2
     \xi \cdot \left[ 1+a\frac{{\vec p_{\text{e}}}
      \cdot{\vec p_\nu}}{E_{\text{e}}
       E_\nu}+b\frac{m_{\text{e}}}{E_{\text{e}}}
     + {\vec{\sigma}}_{\text{n}}\cdot
         \left(A\frac{\vec p_{\text{e}}}{E_{\rm e}} +
           B\frac{{\vec p_\nu}}{E_\nu} + \ldots \right) \right]\,.
     \label{eq:CorrCoeffDefinition}
\end{equation}
Here $\vec{p}_{\text{e}}, \vec{p}_\nu, E_{\text{e}}, E_\nu$, are the
momenta and energies of the electron and neutrino, respectively, $E_0$
is the endpoint energy of the electron, and $\vec{\sigma}_{\text{n}}$ is
the neutron spin.  In the SM, $\xi = G_{F}^2V_{ud}^2 (1+3\lambda^2)$,
and the key decay observables, $a$, the neutrino-electron correlation
coefficient, $A$, the beta asymmetry, and $B$, the neutrino asymmetry,
are fully determined by $\lambda$, while the Fierz term $b\equiv 0$.
Parameters $a$, $A$ and $B$ provide three independent ways to determine
$\lambda$; $A$ and $a$ have comparable sensitivities to $\lambda$, while
$B$ is about 4$\times$ less sensitive.  In practice, $A$ has been
measured much more extensively than $a$, and it dominates the
determination of $\lambda$.

\section{Current status of $V_{ud}$ from neutron $\beta$ decay %, and
%  near-term outlook}
        }

The neutron lifetime is measured by two methods: (a) counting neutron
decays in a fiducial volume containing neutrons (usually from a cold
neutron beam), and comparing the result with the number of neutrons
present in that same volume (the ``beam method''), and (b) storing
ultracold neutrons (UCNs) in a bottle, and observing the decay of the
number of neutrons with time (the ``storage'' or ``bottle'' method).
The bottle may be material or magneto-gravitational.  In the beam
method, the neutron lifetime is measured as $\tau_{\text{n}}=N/r$, where
$N$ is the number of neutrons present in the fiducial volume at any
time, and $r$ is the rate of the detected decays.  Each method comes
with its own set of challenges.  In the beam method both $N$ and $r$
must be measured absolutely.  While the storage method does not require
precise knowledge of the detector efficiency, its primary challenge is
to account accurately for the loss rate of neutrons to processes other
than decay, e.g., capture or up-scattering on walls.  Detailed
discussion of the techniques and systematics for both methods can be
found in recent reviews \cite{Bae14,You14,Wie11}.

\begin{figure}
  \centerline{
   \parbox{0.6\linewidth}{
     \includegraphics[width=\linewidth]{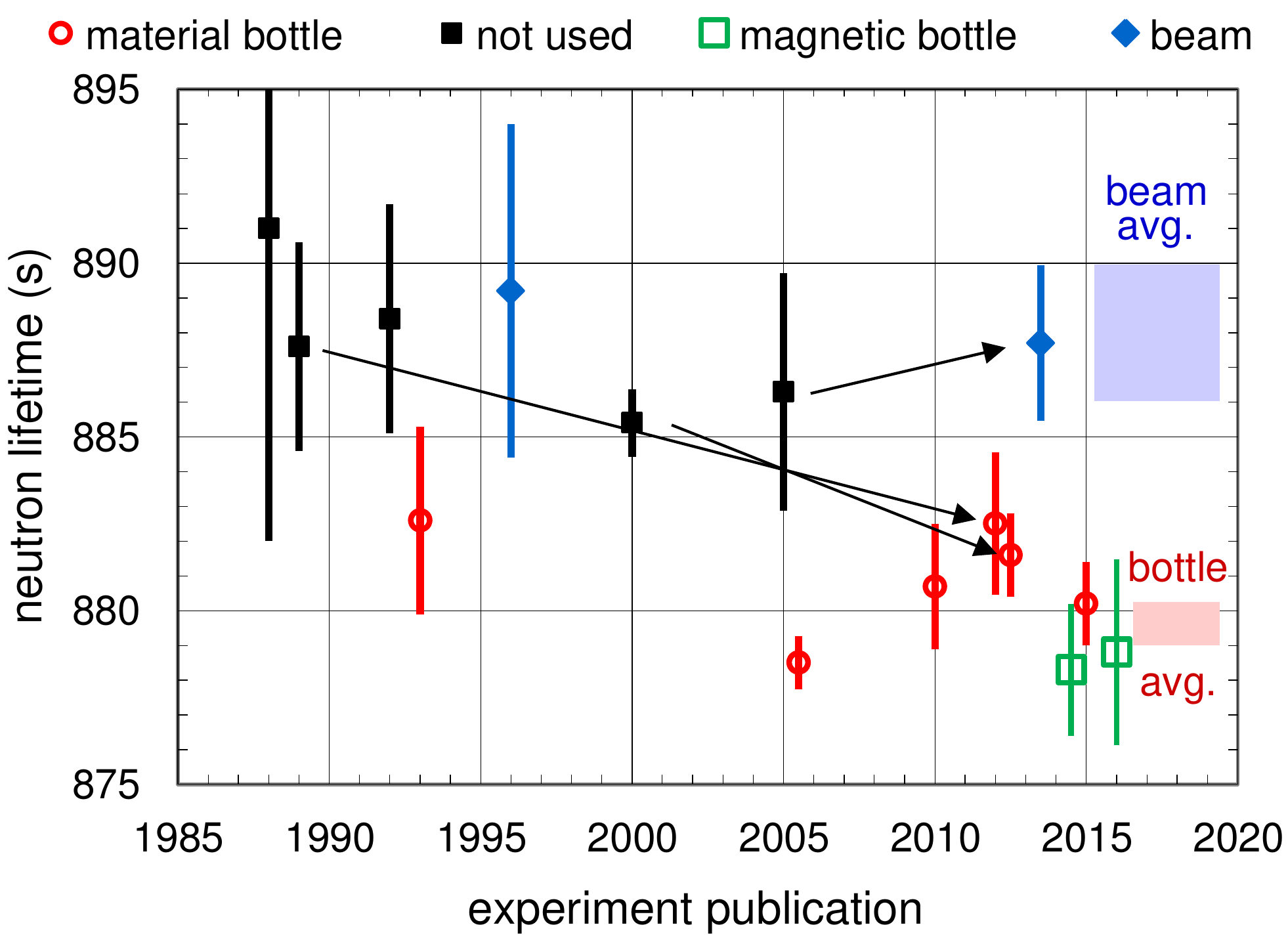}
             }           }
  \caption{Current status of neutron lifetime measurements;
    results are shown according to the year of publication.  Blue
    diamonds: beam method measurements \cite{Byr96, Yue13}.  Red
    circles: material bottle measurements
    \cite{Mam93,Ser05,Pic10,Ste12,Arz12,Arz15}.  Green squares:
    magneto-gravitational trap measurements \cite{Ezh14,Mor16}.  Black
    solid squares: results not used in the global averages, or
    superseded by subsequent publications.  Shaded rectangles on the
    right denote the beam method average, $\tau_{\text{n}}^{\text{beam}}
    =888.0 \pm 2.0$\,s (light blue), and the bottle method average,
    $\tau_{\text{n}}^{\text{bottle}}= 879.6 \pm 0.6$\,s (pink).  The
    discrepancy between the two averages exceeds $4\sigma$.}
  \label{fig:tau_n_2016-11}
\end{figure}

The current status of the neutron lifetime measurements is summarized in 
Figure~\ref{fig:tau_n_2016-11}.
Two striking features of the results emerge: a strong downward shift
over time, brought about by the bottle measurements, and the $>4\sigma$
discrepancy between the storage and beam method experiments.  Both
methods have recently experienced significant improvements, and new
results will be forthcoming.  Key unanswered questions concern: (a) the
ultimate competitiveness of the beam method, (b) whether or not the
above discrepancy will persist, and (c) if so, how the results from the
two methods may be reconciled.  We next turn our attention to $\lambda$.

% \section{Measurements of $\lambda=G_A/G_V$ in neutron decay}

\begin{figure}[b!]
  \centerline{\includegraphics[width=0.45\linewidth]{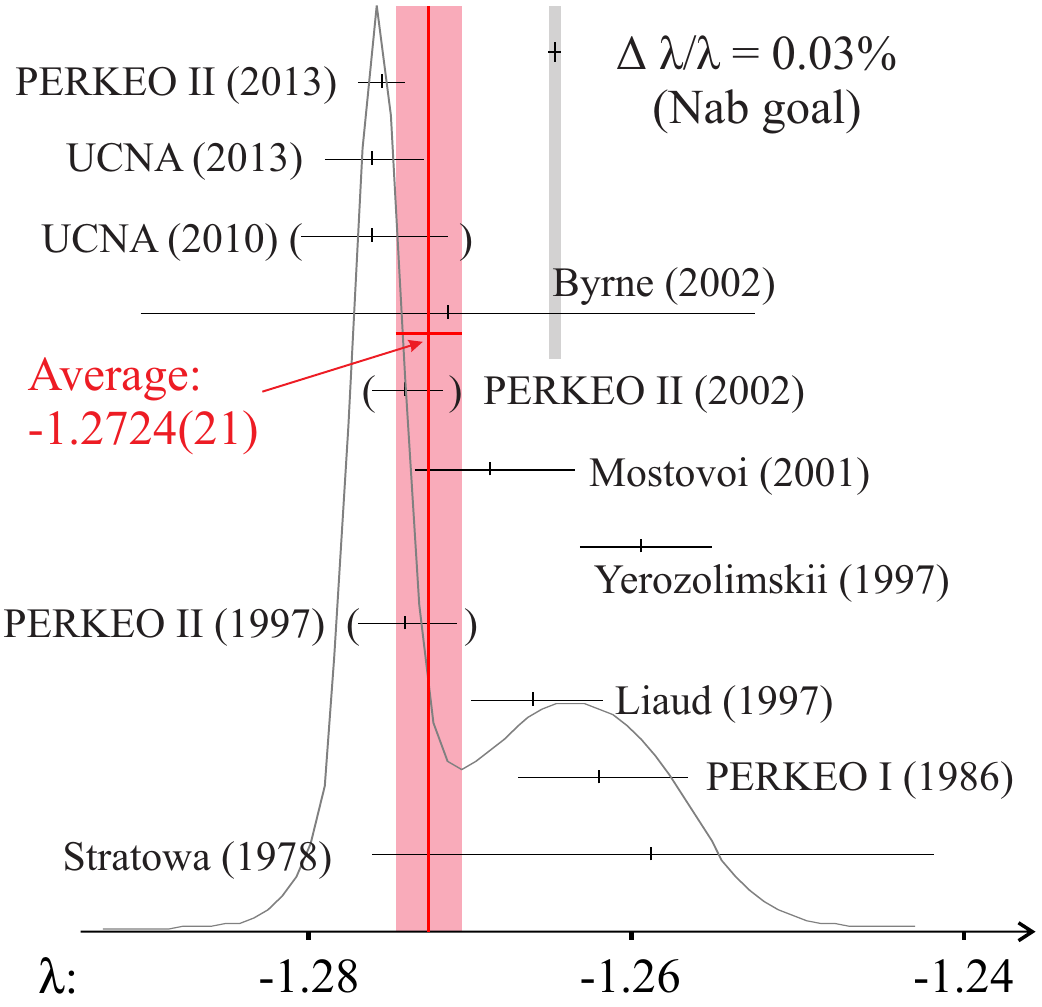} }
  \caption{Current ideogram of the values of $\lambda=G_A/G_V$, with the
    current global average.  For more information see, e.g.,
    \cite{Bae14}.  The planned precision of the upcoming Nab experiment
    \cite{Nab,Bae14} is also indicated.}
  \label{fig:ideog_lambda}
\end{figure}

Figure \ref{fig:ideog_lambda} summarizes the currently available
information on the evaluation of $\lambda=G_A/G_V$.  Most of the data
come from measurements of the $\beta$ asymmetry in neutron decay; the
two measurements of $a$, by Stratowa (1978) \cite{Str78} and Byrne
(2002) \cite{Byr02} are an order of magnitude less precise.  Just as for
$\tau_{\text{n}}$, there appears to be a gradual drift in the measured
values with time, towards greater $|\lambda|$, leading to a poor
confidence level for the global average.

Due to the large remaining uncertainties in $\lambda$, and to a lesser
extent to ambiguities present in the $\tau_{\text{n}}$ data, the
uncertainty in the value of $V_{ud}^{\text{n}} = 0.9758(16)$ exctracted
from neutron decay, is significantly greater than that for SAF nuclear
$\beta$ decays, $V_{ud}^{\text{SAF}} = 0.97417(21)$, which is dominated
by theoretical radiative correction uncertainties.  More work is needed
to make the neutron result competitive with SAF decays, primarily on
$\lambda$, although $\tau_{\text{n}}$ consistency should also improve.

\section{Outlook for the near term} 

Happily, there is a tremendous amount of activity on ongoing and new
precise measurements of the relevant observables in neutron $\beta$
decay.  These experiments are listed in summary form in
Table~\ref{tab:exp_summary}.
\begin{table}
 \centerline{
  \begin{tabular}{ccccl}
   \Hline 
   experiment & observable & goal uncert. & technique & facility/group  \\
   \hline
   BL2  & $\tau$ & 1\,s   & cold $n$ beam & NIST \\ 
   BL3  & $\tau$ & $<0.3$\,s   & cold $n$ beam & NIST \\ 
  JPARC $\tau$  & $\tau$ & $<0.3$\,s   & cold $n$ beam & J-PARC \\ 
  Gravitrap & $\tau$ & 0.2\,s   & UCN/material bottle & PNPI and ILL  \\ 
  Je\v{z}ov & $\tau$ & 0.3\,s   & UCN/magnetic bottle & PNPI and ILL  \\ 
  HOPE  & $\tau$ & 0.5\,s   & UCN/magnetic bottle
                                           & ILL (supertherm.\ source)  \\  
  PENELOPE & $\tau$ & 0.1\,s   & UCN/magnetic bottle & TU Munich  \\ 
  Mainz & $\tau$ & 0.2\,s   & UCN/magnetic bottle & Mainz TRIGA source  \\ 
  UCN$\tau$ & $\tau$ & $\ll$\,1\,s & UCN/magnetic bottle & LANSCE UCN source \\
  \hline
   UCNA & $A$  & 0.2\%   & UCN       & LANSCE UCN source \\
   PERKEO III
        & $A$  & 0.19\%  & cold $n$ beam & MLZ (Munich) and ILL \\ 
   PERC & $A$  & 0.05\%  & cold $n$ beam & Munich \\ 
  aCORN & $a$  & $\sim$1\% & cold $n$ beam & NIST \\ 
  aSPECT& $a$  & $\sim$1\% & cold $n$ beam & Mainz and ILL \\ 
    Nab & $a$  & 0.1\%   & cold $n$ beam & SNS \\ 
   \Hline
  \end{tabular}
            }
  \caption{Ongoing and planned/funded neutron beta decay measurements,
    indicating the observable measured, the goal uncertainty level,
    as well as the general measurement method applied.}
  \label{tab:exp_summary}
\end{table}
The new BL3 and JPARC$\tau$ experiments promise a meaningful comparison
of $\tau_{\text{n}}^{\text{beam}}$ with results from the storage
experiments.  The latter are in full swing, with the emphasis shifting
somewhat toward magneto-gravitational traps.  The much needed
improvement is also forthcoming for $\lambda$, with Nab promising to
bring the precision of $a$ to the level comparable to that of $A$.
There is a healthy mix of experimental techniques both in terms of beam
(cold vs.\ UCN) and spectrometer/detector designs, all of it pointing
toward significant near-term improvements.

\end{document}